\documentstyle[aps,prl,epsf]{revtex}  

\begin{document}
\twocolumn[\hsize\textwidth\columnwidth\hsize\csname
@twocolumnfalse\endcsname

\title{On maturation of crack patterns}
\author{E. A. Jagla}
\address{The Abdus Salam International Centre for Theoretical Physics\\
Strada Costiera 11, (34014) Trieste, Italy}
\maketitle

\begin{abstract}

Superficial (two dimensional) crack patterns appear when a thin layer of material 
elastically attached to a substrate contracts.
We study numerically the maturation process undergone by these
crack patterns when they are allowed
to adapt in order to reduce its energy. The process models the evolution in depth of cracks 
in geological
formations and in starch samples (`columnar jointing'), and also
the time evolution (over thousands of years) of crack patterns in frozen soils.
We observe an evolution towards a polygonal pattern that consist of a fixed distribution
of polygons with mainly five, six and seven sides. They compare very well with known experimental examples.
The evolution of one of these `mature' patterns upon reduction of the degree of contraction 
is also considered. We find that the pattern adapts by closing some 
of the cracks and rearranging those in the immediate neighborhood.
This produces a change of the mean size of the polygons,
but remarkably no changes of the statistical properties of
the pattern. Comparison with the same behavior recently observed in starch samples is presented.

\end{abstract}
%\maketitle
\vskip2pc] \narrowtext

\section{Introduction}

Consider a thin layer of a solid material elastically attached to a substrate. If the material
contracts (or the substrate expands), elastic stresses appear in it. When these stresses
are sufficiently high, cracks can appear in the material, giving rise to a fragmentation process.
Well known examples of this phenomenon are cracking on mud and paints. In these cases the water evaporation
produces the contraction of the material that is responsible for cracking. In other cases, as in
the cracking of ceramic coatings, it is typically the contraction upon cooling that generates the same phenomenon.
Fragmentation is known to produce a two dimensional pattern of cracks whose statistical properties
have been studied theoretically\cite{teor}, 
experimentally\cite{exper} and numerically\cite{numer,numer2}. With some variations
depending on the particular case, these crack patterns are hierarchical structures, with younger 
cracks meeting older ones perpendicularly. Then most crack joints are `T' shaped\cite{nota2}, with the horizontal part
being older than the vertical part.

There is however a small number of remarkable  cases in which fragmentation crack patterns undergo
a `maturation' process. This means that starting from a hierarchical pattern as described above, cracks can
adapt smoothly to optimize its configuration. This optimization process is driven by the tendency of the
crack pattern to reduce its mechanical (elastic plus crack) energy. Special conditions have to 
be fulfilled for this maturation to take place. To modify a given crack pattern, cracks should
be able to displace laterally, and this implies typically the surmounting of enormous energy
barriers (although the final state has lower energy than the original one).
Particular conditions make this lateral displacement possible in (at least)
%\cite{salt}
two different cases.

One is the case of crack patterns formed on the ground of very cold regions of the 
earth\cite{resurfacing}, and also in other planets \cite{marte}. 
In this case the frozen ground (named `permafrost') cracks when the rapidly fallen temperatures
of winter make the surface contract with respect to lower parts of the terrain. This first crack pattern
is of the kind described above. The cracks get filled with new ice and debris, and when temperature rises
after winter
the cracks tend to close. However, the new material that filled the cracks is weaker than the old permafrost,
and the next year cracks open almost on top of the `scars' of first year cracks. However, 
small lateral variations
can occur from one year to the next. There are many reasons that can make a crack to be shifted laterally in one
direction or the other, from one year to the next. Most of these reasons 
(as for instance inhomogeneities in the 
materials) are not expected to bias the shift of the crack in one particular direction. But there is at least
one reason for a crack to shift in a particular direction, and that is the tendency to reduce the energy of
the crack pattern. In fact, from a statistical point of view it is reasonable to expect the crack pattern
to adapt in order to reduce its energy. This tendency provides a bias for the evolution of the crack pattern in
permafrost that over thousands of years is able to qualitatively modify its appearance\cite{resurfacing}.
In fact,  after maturation, crack joints become more `Y' shaped, as this form 
has lower energy than the `T' shaped original joints. 

The second, better known and more remarkable example of crack pattern maturation takes place in
the case of columnar jointing. 
It occurs in basaltic rocks when they cool after its expulsion in a volcanic event\cite{basalto}, and also in desiccating
starch\cite{muller} driven by the shrinkage due to humidity loss. In both realizations, a superficial pattern of
cracks very much like the one described in the first paragraph first develops 
in the material. But here, this crack pattern
penetrates the material as deeper parts of it cool (or dessicate). It is this progression
into the interior that allows the maturation of the pattern to take place, now as a function of
depth,  reaching a polygonal structure whose
further advance defines prismatic columns. In this case there is no true lateral movement of the cracks. But
a description in terms of lateral movement can be given if we choose a reference system
that moves with the penetrating crack front.

We use here a recently developed model of fracture \cite{marconijagla} 
to describe crack patterns in a two dimensional
material elastically coupled to a substrate. In the original formulation of this model cracks have to be 
pinned in some way in order to avoid them to move laterally (since typically this movement is unphysical). Here 
instead, we
take advantage of this movement (driven by the tendency to minimize the energy of the system) to observe
how an originally disordered pattern becomes polygonal during its maturation.
We also investigate the way in which a stable polygonal pattern is modified when the degree of contraction 
is modified.
We observe that some individual cracks disappear (terminate, in the 3D language of columnar jointing) 
when contraction is reduced, 
giving rise to local rearrangements in the pattern.
This mechanism provides a way to change the mean width of the columns as a function of depth 
in the basalt formations and in starch, and it has
been observed to occur in this last case. We finish with a discussion on what the typical width of
columns in three dimensional formations is.

\section{The numerical model}

We use a technique recently developed \cite{marconijagla} 
to treat fracture and cracks in the context of phase field
modeling\cite{phasefields}. The free energy of the system is written in terms of the strain tensor 
$\varepsilon_{ij}\equiv 1/2(\partial u_i/\partial x_j+\partial u_j/\partial x_i)$, with ${\bf u}({\bf r})$ being the local
displacement field.
We choose the form of the free energy in such a way that it 
reduces to the normal elastic energy for small strains, but for large strains it is 
able to describe cracks.
This is achieved by a saturation of the free energy of the system for large values of
$\varepsilon_{ij}$. The inclusion in the free energy of terms proportional to gradients of $\varepsilon$
produces a smoothing of cracks, which although artificial, is however very important 
to us. On one hand it makes the
description isoptropic and 
insensitive to the numerical mesh we use in the calculation (as long as the discretization
is much thinner than the smoothing distance of the fracture). On the other hand it allows
the cracks (that in the regularized theory could be pictorially described as `solitons') 
to move around the system to find configurations of lower energy. This wandering will model the maturation
of the crack pattern. The free energy is taken to be rotationally invariant, in order to describe 
cracks in an isotropic material.

To write down explicitly the equations we actually solved in our two dimensional geometry, 
we first introduce
the following notation for the independent components of $\varepsilon$\cite{kartha,shenoy}
\begin{eqnarray}
e_1&\equiv&(\varepsilon_{11}+\varepsilon_{22})/2\nonumber\\
e_2&\equiv&(\varepsilon_{11}-\varepsilon_{22})/2\\
\label{dos}
e_3&\equiv&\varepsilon_{12}=\varepsilon_{21}\nonumber
\end{eqnarray}
which are named respectively the dilation, deviatoric and shear components.
These three variables 
are not independent. They satisfy the St. Venant compatibility constraint\cite{kartha,shenoy}

\begin{equation}
(\partial^2_x+\partial^2_y)e_1-(\partial^2_x-\partial^2_y)e_2-2\partial_x\partial_y e_3=0.
\end{equation}

The free energy density is
\begin{equation}
F(\varepsilon)=\frac{F^0(\varepsilon)g}{\left[1+F^0(\varepsilon)/f^0\right]}
\label{g}
\end{equation}
where
\begin{equation}
F^0(\varepsilon)= B (e_1-e_1^0)^2 +\mu \left( (e_2-e_2^0)^2+(e_3-e_3^0)^2\right )
\label{g0}
\end{equation}
and $B$ and $\mu$ are related to the two dimensional bulk and shear modulus of the material. $e_i^0({\bf r})$
are externally controlled functions that allow to prescribe the locally preferred state of the system,
and $g({\bf r})$ is another (positive) function that will be used to model
some random inhomogeneities in the system. The limiting value $f_0$ of $F$ for $\varepsilon \rightarrow\infty$ 
(assuming $g=1$), is related to the crack energy in the model.

Regularization of cracks is provided by a gradient term $F_g$ in the free energy density, 
that we choose to be of the form
\begin{equation}
F_g=\sum_{i=1,2,3} \alpha_i (\nabla e_i)^2,
\end{equation}
where we have to choose 
$\alpha_2=\alpha_3$ to retain rotational invariance.

An additional ingredient that has to be added here with 
respect to the basic model of Ref. \cite{marconijagla} is the inclusion of the
elastic energy  density $F_{el}$ of the system attached to the substrate. In terms of the displacement variables 
${\bf u}$, this elastic energy  can be written in the form
\begin{equation}
\int d{\bf r}^2 F_{el}\equiv \gamma \int d{\bf r}^2|{\bf u}({\bf r})|^2
\end{equation}
where $\gamma$ measures the stiffness of the interaction with the substrate.
As we take the components of $\varepsilon$ to be our basic variables, we have to recast this energy in terms
of them. This can be easily done in the Fourier space, and the result is
\begin{equation}
\int d{\bf r}^2 F_{el}=\gamma\int d{\bf k}^2 \frac{|\tilde e_2({\bf k})|^2+|\tilde e_3({\bf k})|^2}{k^2},
\end{equation}
Where $\tilde e_i({\bf k})$ are the Fourier transforms of the original $e_i({\bf r})$.
The equations of motion are taken to be of the overdamped form, namely
\begin{equation}
\frac{\partial e_i({\bf r})}{\partial t}=\lambda \frac{\delta F}{\delta e_i({\bf r})}~~~~~(i=1,2,3)
\end{equation}
where
\begin{equation}
F=\int d{\bf r}^2 \left (F+F_{el}+F_g\right )
\label{f}
\end{equation}
The Saint Venant constraint is implemented by using a Lagrange multiplier.

\section{Results}

We did the simulations on a square mesh of 512$\times$512 elements, using periodic boundary conditions.
Starting from the flat configuration $e_1({\bf r})=e_2({\bf r})=e_3({\bf r})=0$ 
we simulated a uniform and abrupt contraction of the system by taking 
$e_1^0({\bf r})=c$, $e_2^0({\bf r})=0$, $e_1({\bf r})=0$, $c=0.7$. 
We introduce also a finite disorder, taking $g({\bf r})$ in (\ref{g}) to be a random function on the lattice,
uniformly distributed between 0.75 and 1.25. We keep $B$ and the mesh discretization $\delta x$ 
as scale-fixing parameters, and take
$\mu=0.5B$, $\alpha_i=0.5B\delta x^{2}$ ($i=1$, 2, 3),  $\gamma=0.0025 B/\delta x^2$,  and $f_0=0.5B$.
The crack energy per unit length $\eta$ is then $\eta=f_0 \delta x$.
Under these conditions 
we solve the evolution equations. We see in Figs. \ref{f1} and \ref{f2} snapshots of 
the time evolution of the system. The figures are done 
by marking the points in which $e_1>0.5$. According to
our definition of the free energy (\ref{g}) and (\ref{g0}), 
this is a reasonable definition of `broken' elements (results do not depends strongly 
on the threshold value used). In this way 
we are basically
plotting the cracks present in the system. It is important to note that 
due to the finite value of $\alpha_i$, broken
elements do not form strictly one-dimensional `strings' in the system, but they
clusterize, making cracks acquire a finite width, as it is apparent in the pictures. This is a 
crucial point to simulate an isotropic system.

We can distinguish two different stages in the temporal evolution. During the nucleation stage (Fig. \ref{f1})
cracks appear rather disorderly in the
system and propagate around. The pattern that forms is very dependent on many details of the simulation, as for
instance the amount of disorder present. %\cite{otranota}
This 
is the kind of pattern we have described in the introduction as a
fragmentation pattern. During a second stage the maturation of the pattern occurs (Fig. \ref{f2}). 
This is observed as a progressive lateral
displacement of the cracks towards a configuration of lower energy. It is necessary to emphasize again that
in standard fragmentation processes this maturation cannot take place, as cracks are rigidly located in their
positions. 
%Only in the special cases discussed above the maturation can occur. 
In our numerical model cracks can in fact
move laterally, since this does not imply the surmounting of a large energy barrier. 
%It is the same idea as for
%instance in the movement of solitons in the Frenkel-Kontorova model\cite{fk}: they move more easily 
%the more delocalized they are.

The lateral movement of cracks in our model is however rather slow 
compared to its nucleation, and that is why
it is not seen on the timescale of the nucleation stage. 
We stress that we are not forcing the crack
pattern to become polygonal, or cracks to terminate onto other cracks, 
it is the system itself that prefers this kind of configuration as
this reduces its energy. The final, stable pattern is that at the bottom right of Fig. \ref{f2}. 
It corresponds to 
a relative minimum of the energy of the system,
the absolute minimum being a perfect hexagonal pattern with a polygon size (calculated numerically with the
same model) as indicated also on 
Fig. \ref{f2}.
The mature pattern contains mostly polygons of five, six, and seven
sides, and a small number with four, and eight sides. They are 
statistically very similar to those in real columnar formations (see Fig. \ref{f5} below and Fig. 8
in Ref. \cite{jaglarojo}). We note that the mean area of polygons for different number of sides
follows a linear relation, known as the Lewis law, after he encountered it
in other two dimensional patterns\cite{lewis}. This law follows if the pattern is assumed
to be maximally random\cite{rivier}.

The present results can be compared with those 
obtained previously\cite{jaglarojo} using a phenomenological model for the energy of the cracked
material. 
The present approach is however much more general than that in Ref. \cite{jaglarojo}. 
Here, we are not assuming any phenomenological form of the energy as a function of the 
areas of the polygons, the energy of the system builds up from the free energy  
presented in the previous section.
In addition, crack segments are not forced here to be 
straight, and in fact we can see in the last panel of Fig. \ref{f2} that some of them are slightly 
curved. The curvature occurs particularly 
when there is a large difference between the
areas of polygons on both sides of the crack 
segment, always curving it in the direction in which areas tend to
be closer. The reason for this is again energetic: slightly curving a crack does not pay much
crack energy, but produces a gain in elastic energy if the areas of the two adjacent polygons 
tend to become closer to each other. This 
curvature has been in fact observed to occur in a full
three dimensional calculation for a simple geometry\cite{salibajagla}.

An interesting problem to be investigated with the present model is
the way in which a stable polygonal pattern changes when there are changes in the 
parameters that control the extent of contraction. 
As an outcome of this analysis we will get an idea 
of the expected evolution of the patterns down in the columnar formation (after the first maturation), 
since the thermal stresses in deeper parts of the material are lower than close to 
the surface.
Since a lower grade of contraction corresponds to an ideal pattern with larger polygons,  
we may wonder what is the way (if any) in which one of our patterns adapts to the new conditions.
We present in Fig. \ref{f4} the results of simulations when the extent on contraction $c$ is reduced.
We see that there is an increase in the mean area of the polygons when $c$
is reduced. The
area increase is not homogeneous over all polygons, but occurs due to the disappearance of particular 
crack segments, merging two (or three) adjacent polygons into one. After the disappearance of the crack segments
there is a local rearrangement of the pattern which adjusts to the new configurations. 
Those regions in which no crack disappear remain perfectly stable despite the change in the contraction.

The evolution of
the mean area $A$ of the polygons as a function of $c$ is shown 
in Fig. \ref{f5} along with the ideal area $A^{{id}}$ of the hexagons in the
perfect hexagonal pattern of minimum energy. We see that the evolution tends to follow that of the ideal
structure, although $A$ is always smaller than $A^{{id}}$ \cite{nota}.
We note also that in the present model there is a critical value of $c$ ($\sim 0.42$) 
for which $A^{{id}}$ diverges, and we
expect the same occurs for non ideal patterns. This happens
because the elastic energy per unit area gained when generating a polygonal crack pattern decays very rapidly when 
the size of the hexagons increases sufficiently. The sum of this elastic energy plus fracture energy 
may not have a minimum
with respect to the area of polygons if the degree of contraction $c$ is too small. Note that the same does
not apply to a real three dimensional columnar case (see next section).

It is remarkable that the statistical
properties of the pattern do not change appreciable during this relaxation stage. In Fig. \ref{f6}
we see that despite 
a change in the mean area by more than fifty percent,
statistical distribution of polygons by number of sides and areas
remain  constant within numerical fluctuations associated to the finite size of the system\cite{notan}.
Note that it is precisely the `imperfection' of the crack pattern that makes possible the adaptation of the
mean area to a condition of lower contraction. For a perfect hexagonal pattern it is impossible to find 
a way to adapt the pattern slightly and obtain another hexagonal pattern with slightly larger polygon
area. In our case the mean area of the pattern is increased by making some 
crack segments between polygons disappear.

\section{The problem of the typical column width in columnar formations}

The two dimensional model we have studied is perfectly well defined, and provides values for the
size of the polygons in the ideal hexagonal pattern that minimizes the total energy of the system.
We want to comment at this point to what extent these two dimensional results can be
applied to the full three dimensional columnar problem.
For a straightforward application to be possible, the elastic energy stored in a columnarly cracked
three dimensional material should be stored in a layer around the crack front of thickness $w$, in such a
way that $w$ is much smaller than the typical column width $l$ ($l\sim A^{1/2}$). If this is satisfied,
the two dimensional description is directly
applicable. The only consideration to be made is that two-dimensional variables have to be
scaled from the three dimensional variables using $w$. 
For instance, 
the effective crack energy per unit length $\eta$ and elastic constants $B$ and $\mu$
of the two dimensional description are obtained from the real three dimensional values as
$w\eta^{(3d)}$, $wB^{(3d)}$ and $w\mu^{(3d)}$.
Unfortunately the condition $w<< l $ is never satisfied. In fact, 
the elastic energy of a columnar formation is
stored in a portion of thickness $w\gtrsim l$ around the crack front\cite{jagla}. 
The coincidence of the statistical properties of our two
dimensional patterns and
those in true three dimensional cases indicates that 
these properties are robust with respect to this difference. However, the calculation
of the ideal size of the perfect hexagonal pattern (and then an estimation of the typical size of non-perfect
real patterns) has to be reconsidered for the three dimensional case.
In fact, in our two dimensional model, in which a
layer of material is attached to a substrate, there is an ideal hexagonal pattern of well defined polygon size
that minimizes the energy of the system. The application
of the same principle of minimizing the {\em total} energy leads in the three dimensional case to
nonsense: the contribution of the fracture energy to the total energy is always much larger than the elastic
contribution. A minimum can only be obtained with no cracks at all.

The correct
way to pose the problem of the typical size of the polygons in three dimensions is the following: 
in the three dimensional case, we have a temperature profile that we assume to be dependent only on depth $z$,
passing more or less steeply from $T_0$ at $z\rightarrow -\infty$ to 
$T_1>T_0$ at $z\rightarrow \infty$, in such
a way that a temperature front can be defined.
At time $t=0$ the temperature front is assumed to be located at $z=0$.
We will consider the idealized case in which the temperature profile is rigidly 
displaced towards the interior as a function
of time, with some fixed velocity $v$, namely $T(z,t)=\tilde T(z-vt)$. We assume that a stable 
polygonal pattern of fractures has formed, and that its front is located at some depth $z_0$, which moves down
locked to the temperature profile, namely $z_0=z_r+vt$, where $z_r$ measures the relative position of the crack
pattern and the temperature profile. The value of $z_r$ 
depends mainly on the typical size of the pattern $l$ and the overall
temperature difference $\Delta T=T_1-T_0$.
A previous stability analysis has shown in a simplified 
case\cite{jagla} that under the present conditions, patterns
with different $l$ can
propagate in a stable manner, with $z_r$ being a decreasing function of $l$, namely
larger patterns are more retarded with respect to the temperature front.
However there is a limit to this stable propagation. If
$l$ or $\Delta T$ are too small, the crack front becomes unstable: not all cracks can propagate. 
It is tempting to argue (and this is also based upon what is observed in three dimensional starch samples, see
below) than in this case some crack segment will remain halted, and the rest of the pattern
propagates. In this way $l$ is effectively 
increased and the crack front becomes closer to the temperature front, in such a
way that the new pattern is now stable. In a situation in which $\Delta T$ decays smoothly with time
(whereas at the same time the front penetrates the material),
we may expect that the pattern will always be located at the value $z_{cr}$ that marks the 
limit between stable and unstable propagation. 
This is the condition that determines the size
of the columns in terms of the temperature
profile, the elastic properties of the material and the crack energy. 
For the case of a sharp temperature jump and
generalizing the two dimensional expressions for elastic and fracture energy
in Ref. \cite{jagla}, we obtain that the crack front 
is located precisely at the border between stable and unstable
regions when $B (\alpha \Delta T)^2l/\eta$ is some constant value $k$ of order unity (this value is not easy
to calculate). Here $\alpha$ is the thermal expansion coefficient and $B$ is a typical elastic constant of the
material.
From here we obtain the typical width of the columns as
\begin{equation}
l=k\frac{\eta}{B}(\alpha\Delta T)^{-2}
\label{l}
\end{equation}
The typical size is then positively correlated to the crack energy, and negatively correlated with the elastic
stiffness of the material, both facts being qualitatively reasonable. The size $l$ is also proportional to the
negative second power of the
temperature jump responsible for cracking. We should keep in mind however that this result
is valid only for the assumed sharp step form of the temperature profile. 
In other cases we should search for the critical position of the crack front $z_{cr}$ along the lines used in 
Ref. \cite{jagla}.
Note that $\alpha\Delta T$ plays in three dimensions the role of the degree of contraction $c$ in
our simulations. The preceding formula indicates that the typical size of polygons diverges only when 
$\alpha\Delta T\rightarrow 0$, contrary to the critical value of $c$ we found in two dimensions (see Fig. 
\ref{f5}). 
This would indicate that if the driving force for cracking is slowly reduced when going deeply into the material 
(as may occur be due to the higher difficulty to expel heat -or humidity in starch- through the upper material)
the size of the pattern should adapt by increasing their typical size, but it would never stop abruptly.

Recent tomography experiments in starch samples \cite{morris} show that termination
and rearrangement of cracks seem to be in fact the main mechanism by which 
the polygonal pattern evolves in depth. In starch samples the humidity gradients are expected to be reduced
when going deeper into the sample, and that is why the typical width of the columns tends to increase.
However, a quantitative verification of a 
relation like (\ref{l}) (or the equivalent one for a more realistic time dependent temperature or humidity
profile) is not possible at present as it would require the {\em in situ}
determination of the temperature profile under which the cracks form, and not only the observation
{\em a posteriori} of column thickness as a function of depth.

\section{Conclusions}

We have studied numerically the formation and maturation process of a two dimensional crack pattern 
that is allowed to adapt to find configurations of minimum energy. The original cracks
appear in a rather disordered way, but the pattern naturally evolves
towards a polygonal configuration with well defined statistical properties. 
We argue that this maturation process occurs in cracks patterns on the ground of 
arctic regions (permafrost) and effectively in the columnar jointing of basalts 
and starches, as a function of depth. Our model allows also to study the evolution of mature polygonal patterns
when the extent of contraction is reduced. We have found that in this case the pattern adapts
by closing (`terminating' in the three dimensional interpretation) some cracks and rearranging 
those cracks in the immediate neighborhood. This evolution has been recently observed to occur in starch samples.
Although it does not contain all features of the full three dimensional problem, our approach
produces patterns of very good statistical agreement with real ones. The issue of the typical
scale of the three dimensional pattern is beyond the reach of the two dimensional model, and we have provided
for this case a plausible description that relates the typical size of the polygons with the elastic and thermal
properties of the material, and with the details of the temperature profile.

%\section{Acknowledgments}

%Gracias...

\begin{figure}
\epsfxsize=3.3truein
\vbox{\hskip 0.05truein
\epsffile{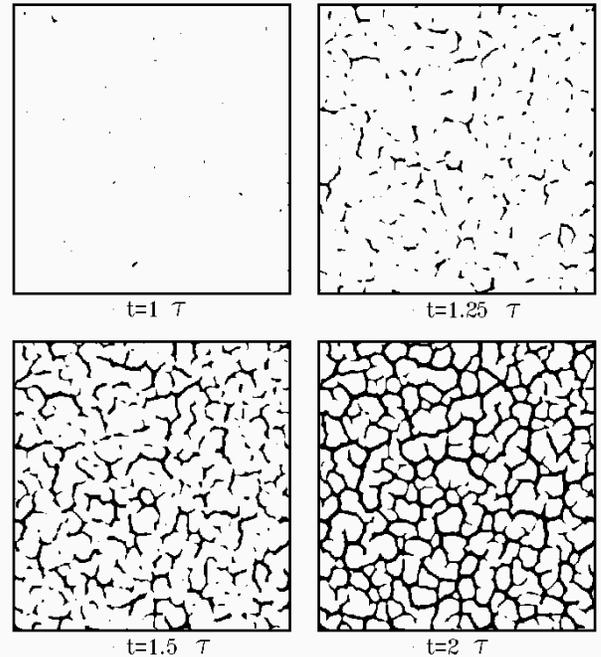}}
\medskip
\caption{Appearance of a typical fragmentation pattern during the first stage of the evolution. 
The time scale $\tau$ is given by $\tau^{-1}=\lambda B $. The contraction imposed is $c=0.7$}
\label{f1}
\end{figure}

\begin{figure}
\epsfxsize=3.3truein
\vbox{\hskip 0.05truein
\epsffile{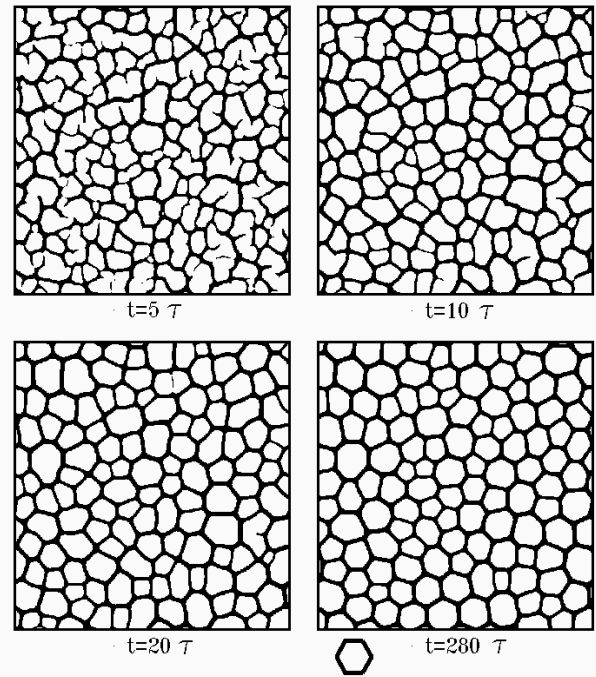}}
\medskip
\caption{Maturation of the fragmentation pattern at longer times (note the change in of time intervals with
respect to previous figure). The lateral displacements of cracks
allows the system to reach a stable state (bottom right) which is a local energy minimum. The size
of the hexagon in the perfect pattern that corresponds to the absolute minimum of the energy is indicated.}
\label{f2}
\end{figure}

\begin{figure}
\epsfxsize=3.3truein
\vbox{\hskip 0.05truein
\epsffile{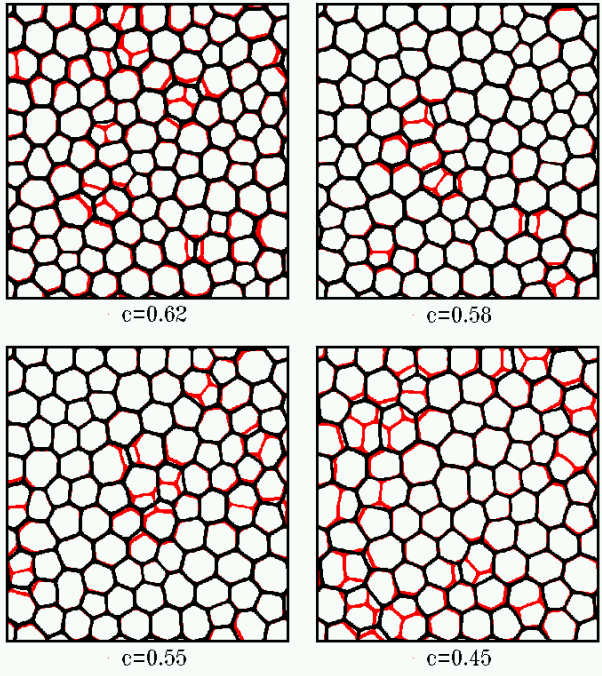}}
\medskip
\caption{Evolution of the mature pattern of Fig. 
{\protect \ref{f2}} (bottom right) upon reduction of the extent of contraction $c$. Note the disappearance of
some cracks and the local rearrangement that occur. To facilitate the visualization 
we plot also the immediately previous pattern. For the first panel the previous pattern is the last one in Fig.
\ref{f2}.}
\label{f4}
\end{figure}

\begin{figure}
\epsfxsize=3.3truein
\vbox{\hskip 0.05truein
\epsffile{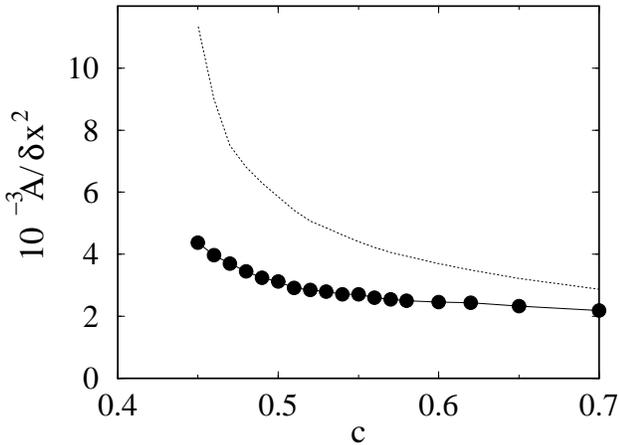}}
\medskip
\caption{Evolution of the mean area $A$ of the polygons as a function of the degree of contraction $c$, when $c$
is reduced from larger to smaller values, and ideal value $A^{{id}}$ of the area of polygons in the perfect
hexagonal pattern that minimizes the energy of the system. There is a critical value of contraction
($c\sim 0.42$) below which the uncracked configuration is the one with minimum energy.}
\label{f5}
\end{figure}

\begin{figure}
\epsfxsize=3.3truein
\vbox{\hskip 0.05truein
\epsffile{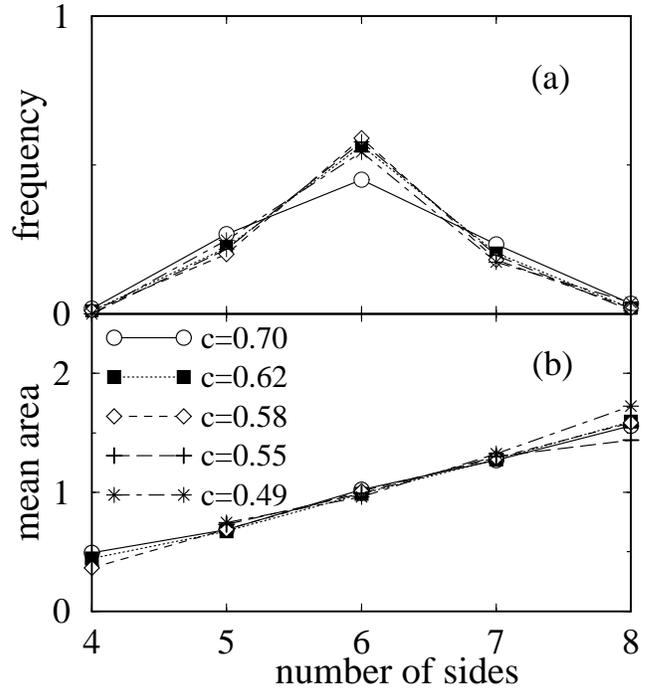}}
\epsfxsize=3.3truein
\caption{(a) Frequency of polygons with different number of sides and (b) mean area of polygons with 
different number of sides (normalized to the mean area of all polygons) for patterns obtained by reducing 
$c$.}
\label{f6}
\end{figure}


\begin{references}
\bibitem{nota}This difference 
can be reasonably considered to be a consequence of the metastabilities in the process we are simulating. 
In fact, if at some
point we revert the evolution of contraction, making now $c$ increase, we have 
observed the nucleation of new crack
segments, typically bisecting the largest polygons. In this evolution the values of $A$ 
we get are {\em larger} then $A^{{id}}$, namely the evolution of $A$ as a function of $c$ has a strong
hysteresis that encloses the ideal value corresponding to the ideal energy minimum.
\bibitem{notan} In Fig. 5(a), it seems that the first stable pattern 
(with the original value of contraction $c=0.7$) is slightly different to the rest, having less polygons
with six, and more with five and 7 sides. We cannot tell at this point if this is systematic or associated to
statistical fluctuations.
\end{references}
\end{document}